\documentclass[%
prl,
twocolumn
]{revtex4-2}

\usepackage{graphicx}
\usepackage{dcolumn}
\usepackage{bm}
\usepackage[T1]{fontenc}
\usepackage[latin9]{inputenc}
\usepackage{natbib}
\usepackage{color}
\usepackage{babel}
\usepackage{amsmath}
\usepackage{amssymb}
\usepackage{graphicx}
\makeatletter
\@ifundefined{textcolor}{}
{%
 \definecolor{BLACK}{gray}{0}
 \definecolor{WHITE}{gray}{1}
 \definecolor{RED}{rgb}{1,0,0}
 \definecolor{GREEN}{rgb}{0,1,0}
 \definecolor{BLUE}{rgb}{0,0,1}
 \definecolor{CYAN}{cmyk}{1,0,0,0}
 \definecolor{MAGENTA}{cmyk}{0,1,0,0}
 \definecolor{YELLOW}{cmyk}{0,0,1,0}
}

\makeatother

\begin{document}
\title{Real-space model for activated processes in rejuvenation and memory behavior of glassy systems}
\author{Mahajabin Rahman and Stefan Boettcher}
\affiliation{Department of Physics, Emory University, Atlanta, GA 30322, USA}

\begin{abstract}
We offer an alternative real-space description, based purely on 
activated processes, for the understanding of relaxation 
dynamics in hierarchical landscapes. To this end, we use the 
cluster model, a coarse-grained lattice model of a jammed 
system, to analyze rejuvenation and memory effects during aging 
after a hard quench. In this model, neighboring particles on a 
lattice aggregate through local interactions into clusters that 
fragment with a probability based on their size. Despite the 
simplicity of the cluster model, it has been shown to reproduce 
salient observables of the aging dynamics in colloidal systems, 
such as those accounting for particle mobility and 
displacements. Here, we probe the model for more complex quench 
protocols and show that it exhibits rejuvenation and 
memory effects similar to those attributed to the complex hierarchical structure 
of a glassy energy landscape. 
\end{abstract}
\maketitle

\section{Introduction}
\label{Intro}
The morphology of energy landscapes in high-dimensional configuration spaces is at the heart of complex dynamics for a broad range of statistical
systems~\citep{Frauenfelder96,Wales03}. Examples are legion in disparate
systems: Glassy materials like amorphous fluids~\citep{Vollmayr-Lee16},
jammed grains~\citep{Liao2019,GB20}, colloids~\citep{Charbonneau13,ditBiot20},
disordered magnets~\citep{Sibani89,Boettcher20c}, crumpling sheets~\citep{Shohat23,Lahini23}, or entangled 
polymers~\citep{Struik78,angell:95,Parker03,roth2016polymer}, all face frustration
while relaxing their free energy. Due to competing variables exerting geometric or energetic constraints on each other,  a complex, multimodal landscape is imposed on the space of
all possible configurations. A universal framework of how a system traverses its landscape can elucidate its collective states, and the transitions between them that result in different phases of behavior. Such a framework is therefore of wide ranging scientific interest.

A standard approach to gain insight into the complexity of the landscape of 
a glassy system, whether in experiment or in simulation, is through a hard 
quench~\citep{Struik78} from the liquid-like high-temperature (or 
low-density) to a low temperature (or high density) regime, initiating a 
non-equilibrium relaxation dynamics known as 
aging~\citep{Lundgren83,Bouchaud92,Cugliandolo93,Hutchinson1995,Jonsson1995,komori99,Kob00b,Rodriguez03,Barrat03,Elmasri05,Yunker09,Amir2012,Vollmayr-Lee16,Bernaschi20,GB20}.
Such a quench takes the system
instantly deep into the glassy landscape. There, a hierarchy of barriers
emerges that quite naturally calls for an effective description
of the ensuing dynamics in terms of a sequence of activated events
that is called \emph{record dynamics} (RD)~\citep{Anderson04}, since that
hierarchy renders all but the largest fluctuations ineffectual and
relaxation is characterized by timescales for barrier crossings that
exceed all others~\citep{Sibani05,SJ13,Boettcher20c,BGS21}. There is significant
experimental evidence indicating the dominance of such large, intermittent
events in driving the relaxation 
dynamics~\citep{Bissig03,Buisson03,Buisson03a,Sibani06a,Yunker09,Kajiya13,Shohat23,Lahini23}. Alternative approaches to describe aging in terms of intermittent events 

As discussed in Ref.~\citep{Robe16}, RD derives its generality from its small set of assumptions about the properties of the energy landscape for a generic glassy system. Energy landscapes are a widely applicable concept across many areas in science and engineering~\citep{Wales03}, describing the configuration space of systems with a large number of degrees of freedom. For RD within that concept, we merely need to stipulate (1) that a complex energy landscape has a rapidly (say, exponentially) growing number of meta-stable states for increasing energy and (2) that lower-energy meta-stable states are more stable, i.e., have higher energy barriers against escape, than those at higher energy. Then, a hard quench entrenches the system with high probability in one of the far more prolific meta-stable states of higher energy, which the system explores through (reversible) quasi-equilibrium fluctuations. Only progressively larger (and increasingly rare), record-sized fluctuation events allow the system to overcome ever larger barriers to tumble irreversibly into the next, marginally more stable local energy minimum.

Accordingly, in RD incremental relaxation is coarse-grained into a sequence of record 
barrier crossing events that are required to unlock farther reaches
in the landscape~\citep{BoSi,BGS21}. 
These records drive the dynamics (i.e., ``set the clock'')
in disordered materials, generically, reminiscent of the concept of "material time"~\citep{PeredoOrtiz22,Boehmer24}.  This ``clock'' for records decelerates at a rate $\lambda(t)\propto1/t$,
as new records are ever harder to achieve. Dynamics proceeds
homogeneously in $\log t$ instead of in linear time~\citep{SJ13,Boettcher18b}, as observed in many experiments for polymers~\citep{Pastore20,roth2016polymer}, colloids~\citep{Courtland03,Yunker09,Pastore22}, granular piles~\citep{BenNaim98,GB20}, or crumpling sheets~\cite{Shohat23}, obtaining for the accumulation of events $\langle n(t,t_w)\rangle\sim\int_{t_w}^t \lambda(\tau)d\tau\sim\log(t/t_w)$. Then, any two-time correlations
become \emph{subordinate}~\citep{Sibani06a} to this clock: $C\left(t,t_{w}\right)=C\left[n(t,t_w)\right]=C\left(t/t_{w}\right)$ for times $0<t_w<t$ after the quench, as shown, e.g., in Ref.~\citep{Robe16}.

As a real-space incarnation of RD, a simple on-lattice ``cluster
model'' has been designed~\citep{Becker14} that captures the combined
temporal \emph{and} spatial heterogeneity found in a generic aging
system. Despite it simplicity, the model has already been shown to 
reproduce~\citep{Robe16,Robe18} salient experimental~\citep{Yunker09} and
simulational~\citep{ElMasri10} results for quenches in colloids.

Further subjecting an aging system to a protocol of temperature shifts should trigger 
rejuvenation and memory effects. In these protocols, the aging process restarts after a second 
quench (rejuvenation), and resumes the dynamics prior to the second quench upon reheating, thus having memory.  In spin glasses, it is easy to 
demonstrate rejuvenation and the imprinting of entire 
histories~\citep{nordblad:98,Dupuis2001,Sibani04a,Dupuis2005,Vincent06,Krzakala2006}
under small variations of temperature and fields after a quench. Similarly, in polymeric and colloidal systems, memory and rejuvenation effects have been known for a long time~\citep{Kovacs63,Di11,Peng16}. Recently, using extensive MD simulations of a structural (colloidal) glass, it was argued that these effects validate mean-field predictions from spin glass theory~\citep{Scalliet19}. 
Here, following the protocol of Ref.~\citep{Scalliet19}, we demonstrate similar rejuvenation and memory effects in the cluster model of RD. Reproducing almost the entire phenomenology in such a minimalistic setting highlights the role of rare activated processes, which are the only driving mechanism in the cluster model. Merely the lack of a realistic equilibrium state in this model leads to unphysical behavior at infinitely long times.

\section{Methods}
\label{Methods}
\subsection{Cluster Model}
\label{ClusterModel}
In the cluster model~\citep{Becker14}, $N=L^d$ particles completely fill a (hyper-cubic) lattice of base-length $L$ in dimension $d$, one on each site at all times. Yet, each particle by itself is either isolated and forms a cluster of size $h=1$, or it is jammed
in with adjacent particles as a member of a cluster of 
size $h>1$. Isolated particles ($h=1$) possess independent mobility, those in clusters with $h>1$ are locked in and require activation to become mobile. At the time of the quench, $t=0$, all particles are mobile, 
owing to the prior ``liquid'' high-$T$ or low-density state of the system. 
Each update step $s$, one randomly chosen site is picked for an update. (Time is measured in units of lattice sweeps, $t=s/N$.) There are two possible outcomes, depending on the state of the particle on that site: 
\begin{enumerate}
\item A mobile particle interacts with a randomly chosen
neighbor and both exchange position, the basic unit of mobility in the model. Whether that neighbor itself was mobile ($h=1$) or already part of a larger cluster ($h>1$), the addition of the mobile particles now leads to a (jammed and thus immobile) cluster with $h^\prime=h+1>1$. 
\item An immobile particle jammed inside a cluster of size $h>1$ may activate a barrier-crossing event with an $h$-dependent probability per sweep~\citep{BoSi09},
\begin{equation}
P(h)\propto e^{-\beta h}.\label{eq:Ph}
\end{equation}
If it occurs, such an event will break the cluster and create $h$ newly mobilized particles.
\end{enumerate}
Thus, following a quench out of the initial liquid
state of mobile particles, clusters form and break up \emph{irreversibly} to re-mobilize and re-distribute their particles to neighboring clusters. For a sufficiently large value of the external control parameter $\beta$ 
(that acts as a density or an inverse temperature), a large fraction of particles soon 
accrete into jammed clusters that only intermittently break up and almost instantaneously feed their particles into ever fewer -- and thus ever larger -- neighboring clusters, which in turn necessitate ever 
larger and thus ever more rare fluctuations, requiring a time \emph{exponential}~\citep{Shore92,Boettcher20c} in 
the size of those clusters. The effect of all regular fluctuations that only
rarely achieve such a significant event beyond reversible in-cage rattle
is coarse-grained into $P(h)$ in Eq.~(\ref{eq:Ph}). Cluster growth ultimately decelerates the dynamics, since only larger and fewer clusters 
remain, which signifies the slow structural changes that characterize aging~\citep{Yunker09,Shohat23}. Note that high or low "density" in this model is dictated via the choice of the temperature-like parameter $\beta$ in Eq. (\ref{eq:Ph}), not by the actual (and always uniform) filling of the lattice. 

\begin{figure}
\centering{}\includegraphics[width=0.85\columnwidth]{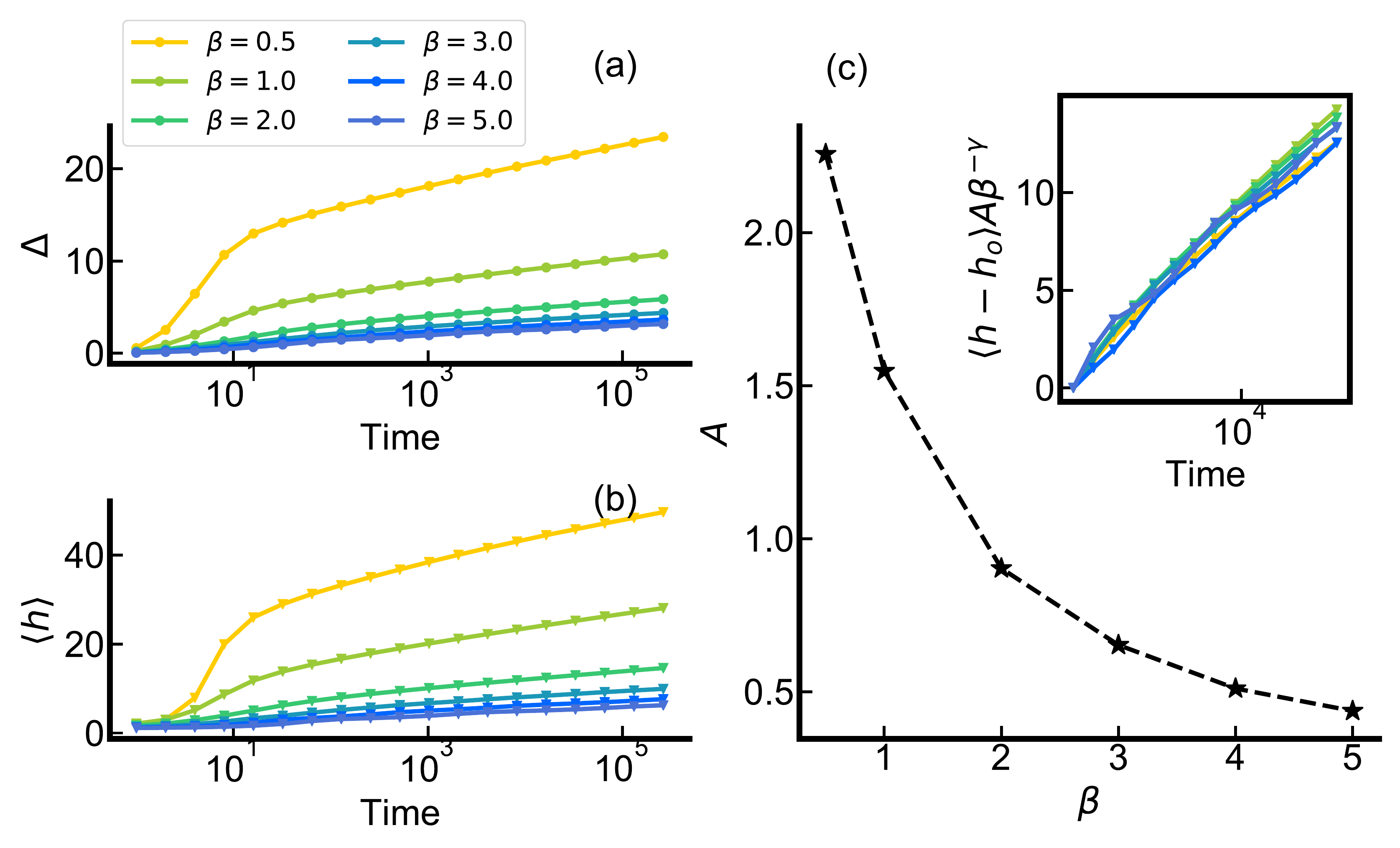}
\caption{\label{fig:beta-comparison} Increase of (a) MSD and (b) average 
cluster size with the logarithm of time $t$ for $t_w=1$, for different values of $\beta$ in 
Eq.~(\ref{eq:Ph}). The results show that motion slows systematically with 
increasing $\beta$. At the shortest times, fast transient effects resulting from the quench predominate, leading to an instant jump $\Delta_0$ in MSD or $h_0$ in cluster size, before the logarithmic scaling sets in.
Panel (c) shows the dependence of the log-slope $A$ on 
$\beta$ in fitting $\Delta\sim A\ln(t)$ to the cluster sizes $\langle 
h\rangle$ in (b), yielding $A\approx \beta^{-\gamma}$ with $\gamma\approx 
0.7$. The inset demonstrates the collapse of the appropriately rescaled 
data from (b).
}
\end{figure}

In Ref.~\citenum{Becker14},
the two-time mean-square displacement (MSD), 
\begin{equation}
\Delta\left(t,t_{w}\right)=\frac{1}{N}\sum_{i=1}^{N}\left\langle \left|\vec{r}_{i}
\left(t\right)-\vec{r}_{i}\left(t_{w}\right)\right|^{2}\right\rangle ,\label{eq:MSD}
\end{equation}
was shown to grow logarithmic as $\Delta\left(t,t_{w}\right)\sim 
A\ln\left(t/t_{w}\right)$, depending on the waiting time $t_{w}$ after the 
quench when the measurement commences. In RD, this is a direct consequence 
of the $\sim A/t$ decline~\citep{Arnold08} in the rate of cluster break-up 
events (such a rate for irreversible events was explicitly verified in 
experimental data for aging colloids~\citep{Robe16}). 
In Fig.~\ref{fig:beta-comparison}, we demonstrate that the proportionality 
factor $A$ is a function of $\beta$, similar to what has been observed
for domain growth in spin-glass simulations~\citep{Sibani18}, granular compaction~\citep{GB20}, but
also for MSD in colloidal experiments at different densities~\citep{Robe16}.

\subsection{Waiting-Time Method}
\label{WTM}
In the following, we use two-step quench temperature cycles in the
cluster model to elicit rejuvenation and memory effects. To that end,
we will simulate the model on a two-dimensional lattice of base length $L=64$ throughout,
employing periodic boundary conditions. Such investigations require
measurements over many decades in time, which is conveniently accomplished
in the cluster model using a waiting-time algorithm~\citep{Dall01,Becker14}.
This method orders all possible events in a chronology
based on their probability of occurring in Eq.~(\ref{eq:Ph}), thus avoiding rejected moves
that occur in conventional Monte Carlo algorithms. Given Eq.~(\ref{eq:Ph}), newly freed particles will join clusters on sub-sweep time scales, meaning that each cluster 
break-up event will typically move many particles nearly simultaneously. That said, this method 
can rapidly telescope into the future within just a few update steps when break-ups become rare. 
Specifically, each of the $k$ clusters is assigned a survival time, 
$\left\{ \delta \, t_{i}\right\} _{i=1}^{k}$, based on its current size 
$h_i$ according to~\citep{Dall01} 
\begin{equation}
\delta \, t_{i}=-\log(X_{i})/ P(h_{i}),\label{eq:WTM}
\end{equation}
with $P(h_{i})$ as given in Eq.~(\ref{eq:Ph}) and made stochastic by 
employing a random number $X_{i}$ sampled from a uniform distribution. 
Then, the event with the lowest 
$\delta t_{\text{min}}=\min_{i}\left\{ \delta t_{i}\right\}$
is selected, updating the global time to $t +\delta t_{min}$
and assigning a new  $\delta t$ to the most recently modified clusters.

\begin{figure*}
\begin{centering}
\includegraphics[width=150mm]{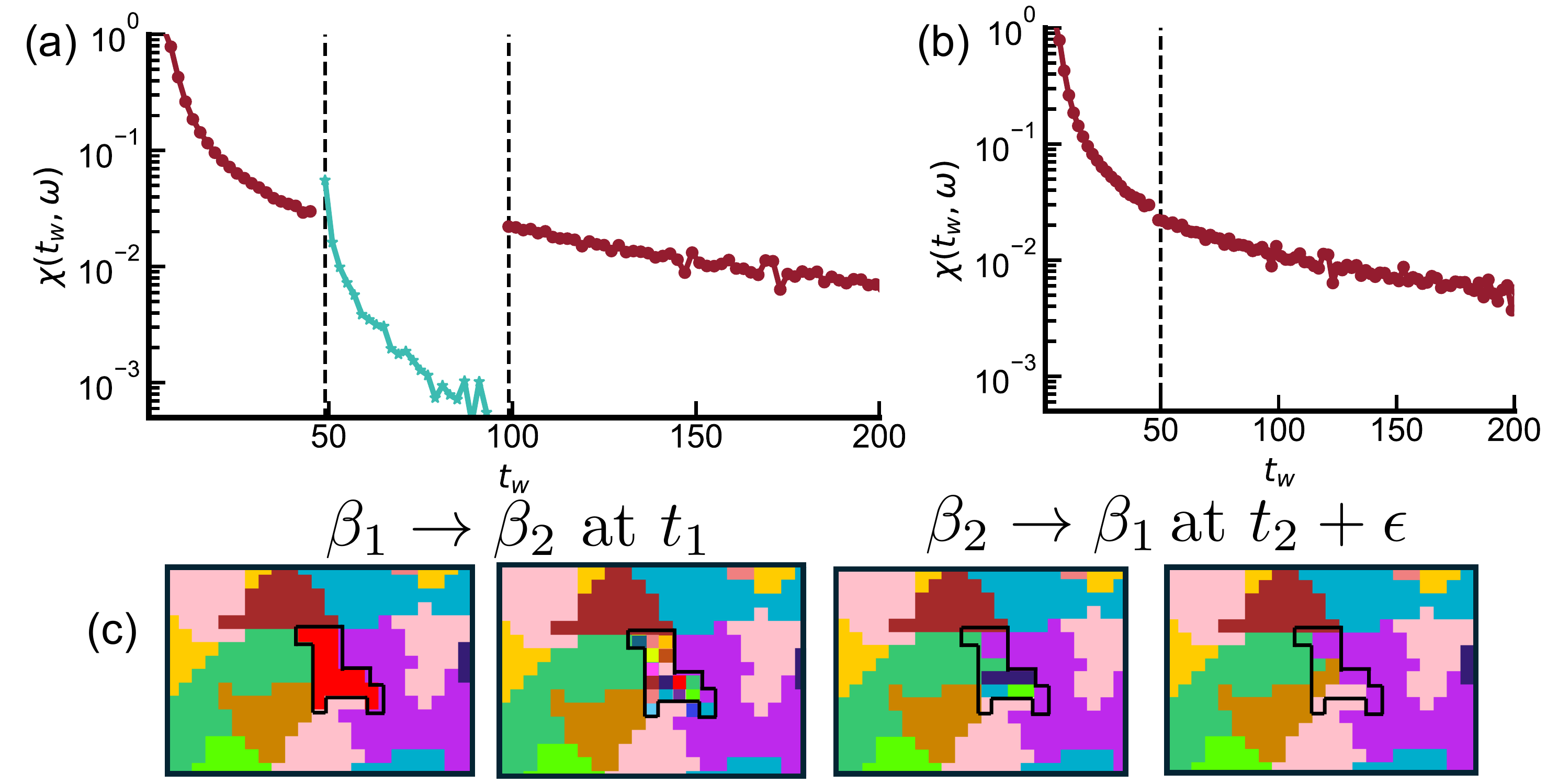}\caption{\label{fig:Rejuvenation-and-memory}
Rejuvenation and memory effects produced on an  $L=64$ square lattice 
subject to a temperature cycle. The system at $t_w=0$ undergoes a hard 
quench to $\beta_{1} = 0.5$, ages until time (in sweeps) $t_{1} = 50$, when 
temperature is reduced once more to $\beta_{2} = 5$. After aging further 
until $t_{2} = 100$, it is reset to $\beta_{1}$. In (a), the susceptibility 
$\chi$ defined in Eq.(\ref{eq:chi}) is plotted as a function of $t_w$ using 
$\tau=\omega^{-1}=2\ll t_1$. In turn, (b) shows that $\chi$, when reheated 
at $t_{2}$, is a continuation of the dynamics from the system prior to the 
second quench at $t_{1}$.  Both can therefore be ``stitched together''. In 
(c), a physical depiction of the situation is provided. This row shows 
the cluster formation by zooming in on a small part of the lattice (different colors indicate distinct 
clusters). The region most affected by the quench at $t_1$ is outlined in all 
the snapshots. There, some cluster of size $h=18$ happens to be in the process of breaking up. 
Solely its freed particles are able to move during a time window of size 
$\tau=2$ after $t_{1}$. A few of them attach to neighboring clusters, the remaining ones form small clusters that can survive for a long time at $\beta_2=5$. When the lattice is reheated to $\beta_1$ at time $t_2$, those small cluster almost instantly (i.e., in a small time interval $\epsilon\ll t_2$) break up and their particles integrate into the surrounding clusters, as they would have done without the second quench at $t_1$. Thus, 
 the cluster-size distribution at $t_1$ is virtually identical to that at $t_2$, once reheated, which is why the dynamics in panel (b) appears to
pick up where they left off prior to the second quench.
}
\end{centering}
\end{figure*}

\section{Results and Discussion}
\label{results}
In Fig.~\ref{fig:Rejuvenation-and-memory}, we illustrate that our
simple model is capable of exhibiting both rejuvenation and memory
effects. Following Ref.~[\citenum{Scalliet19}], we employ the MSD
given in Eq.~(\ref{eq:MSD}) to define a dynamic susceptibility function
\begin{equation}
\chi\left(t_w,\omega\right)=\beta\Delta\left(t_w+\omega^{-1},t_w \right),
\label{eq:chi}
\end{equation}
where $\tau=\omega^{-1}$ sets a time-window over which the decay of the 
instantaneous mobility at time $t_w$ is assessed. The initial quench occurs 
from an infinite temperature ($\beta=0$) to $\beta_{1}=0.5$. At that point, 
$\chi$ drops as a function of $t_{w}$, using a window size of $\tau=2$ (in 
sweeps), while the system is aged up to $t_{1}=50$ sweeps.  At that time 
the system has developed a Poissonian cluster-size distribution with 
average cluster size reaching about $\left\langle h\right\rangle 
\approx30$, see Fig.~\ref{fig:beta-comparison}(b), leaving a number of the 
smallest and most marginally stable clusters below that size most likely to 
break. In fact, as illustrated in 
Fig.~\ref{fig:Rejuvenation-and-memory}(c), a fraction of those clusters are 
in the process of steadily collapsing at time $t_1$ in a large enough 
system. (Since clusters, as well as correlations between clusters, only grow very slowly, there is little difference in averaging over the evolution of one large or many smaller systems, as long as $\langle h\rangle^{1/d}\ll L$.) 

At time $t_{1}$, we perform a further quench of the system, down to 
$\beta_{2}=5$. At this much reduced temperature, only clusters below the 
corresponding average cluster of $\left\langle h\right\rangle \approx3$, as 
taken from Fig.~\ref{fig:beta-comparison}(b), would qualify as unstable on 
this time-scale, i.e., susceptible to breaking up in a time of the order of $t_1$. Clearly, all of the existing clusters are much too large 
and are completely frozen at this temperature. Only those currently freed 
particles from the cluster break-ups  can 
contribute to the instantaneous mobility in this part of the temperature 
cycle. This small but extensive fraction of mobile particles, in turn, 
relives the entire history of an aging system freshly quenched to 
$\beta_{2}$, within the background of otherwise frozen clusters. As in 
Ref.~[\citenum{Scalliet19}], the overall reduction in mobility $\Delta$ is 
partially compensated by the relative factor of $\beta$  in the definition of $\chi$ in Eq.~(\ref{eq:chi}): Here, we have $\beta_2/
\beta_1=10$, while it is $\approx70$ in Ref.~[\citenum{Scalliet19}]. Thus, 
$\chi$ ``rejuvenates'', immediately jumping up above the previous level 
reached before $t_{1}$, before decaying itself. When the temperature is 
then reset to $\beta_{1}=0.5$ after $t_1+t_{2}=100$ sweeps, the impact left 
by the rejuvenating sub-system had a minimal effect on the entire system. 
Merely those clusters in the process of breaking up at $t_1$ \emph{already} have advanced minutely. 
Accordingly, its instantaneous mobility returns to the level frozen in at 
$t_{1}$.

\begin{figure}
\begin{centering}
\includegraphics[width=0.99\columnwidth]{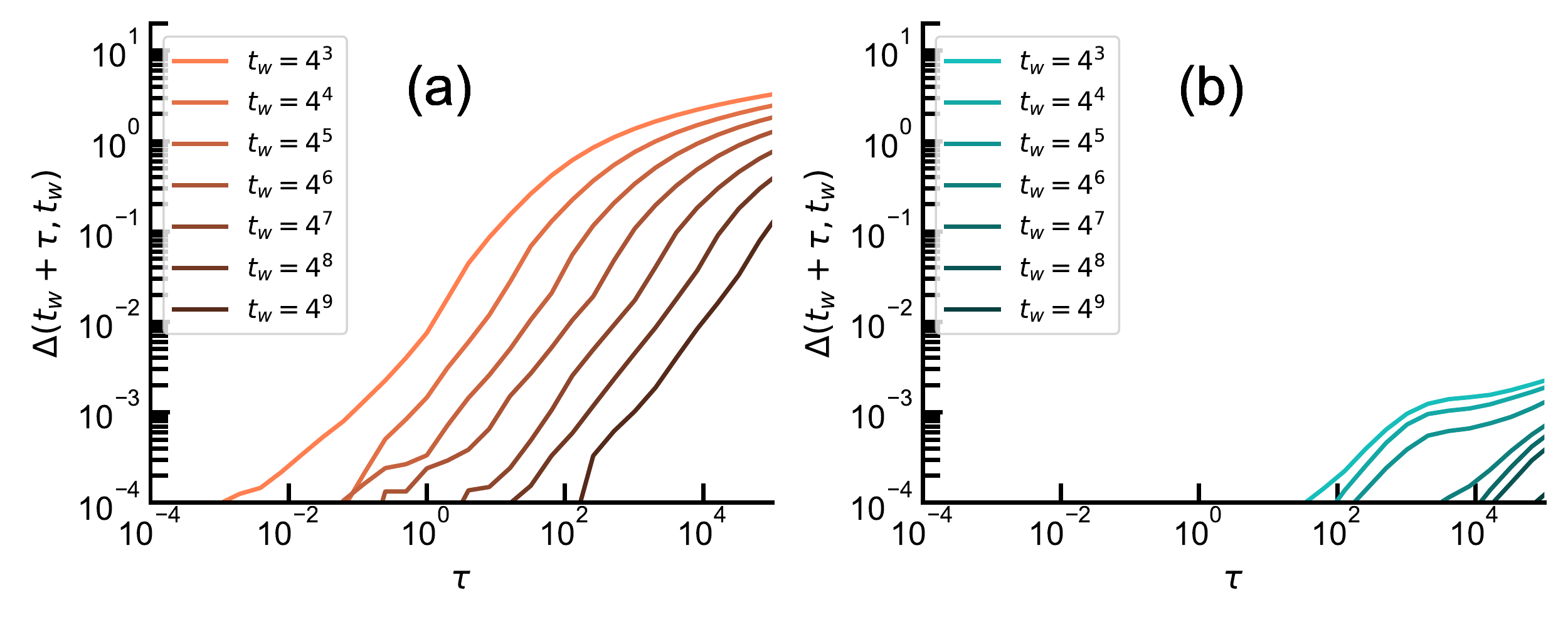}
\par\end{centering}
\caption{\label{fig:MSD-T1}MSD (a) after quench to $\beta_{1}$ = 0.5 
(without subsequent temperature changes) and (b) after a subsequent quench 
at $t_1=50$ to $\beta_{2}$ = 5.0. In both cases, the quenched system is 
aged up to certain waiting time $t_{w}$, before the dynamics of the 
particles are measured relative to the configuration at $t_{w}$ as a 
function of lag-time $\tau=t-t_w$. Both (a) and (b) show the characteristic 
dependence of MSD on the age $t_w$. For (b) this implies that the second 
quench actually rejuvenated the system, albeit at a much lower mobility due 
to the lower temperature. Note that large fluctuations which ensue due to a rare event account for the large gap between MSD curves measured at $t_{w} = 4^5$ and $t_{w} = 4^6$.}
\end{figure}

As a further validation of rejuvenation, Ref.~[\citenum{Scalliet19}] compared the age-dependent 
(two-time) MSD observed following the initial quench to $\beta_1$ with the 
MSD found after the second quench to $\beta_2$ while using its starting point $t_1$ 
as the new origin of time. Indeed, in their Figs.~3 and~4(a), they demonstrate that in 
both measurements the two-time MSD behaves analogously, as if $t_1$ was an entirely 
independent quench.

To replicate these results in RD, we employ for the cluster model the same setting as in 
Fig.~\ref{fig:Rejuvenation-and-memory} but with a simple quench to $\beta=0.5$. Now, 
the system is aged (without second quench) up to various waiting times $t_w$ to 
measure MSD $\Delta\left(t_w+\tau,t_w\right)$ for 
the lag-time $\tau = t - t_{w}$. This data is plotted in Fig.~\ref{fig:MSD-T1}(a), which 
reproduces Fig.~3 of Ref.~[\citenum{Scalliet19}].  It 
demonstrates that a system that was aged up to a time $t_w$ remains confined for a 
corresponding time $\propto t_w$ before exhibiting any discernible MSD. Incidentally, 
this fact, as well as a collapse of this data as function of $t/t_w$, was previously 
explained for experiments on colloids in terms of RD in Ref.~[\citenum{Robe16}].  
Although mean-field arguments would suggest that MSD after a transient should 
saturate at long times~\citep{Jin22}, the 
existence of activated dynamics in real systems induce further (logarithmic) growth. 

More importantly, the rejuvenation effect seen in Fig.~4(a) of Ref.~[\citenum{Scalliet19}] 
is captured for the cluster model in Figure \ref{fig:MSD-T1}(b) which presents the two-time 
MSD of particles for several $t_{w}$ during the second stage of the temperature 
cycle. Having undergone the initial quench to $\beta = 0.5$, the dynamics are evolved up 
to time $t_{1} = 50$ sweeps, at which time the system is cooled down even further to 
$\beta = 5$. Once the particles are quenched to the second temperature, they are aged up 
to a given waiting time $t_{w}$, now taking $t_1$ as the new origin of time. As above, 
the dynamics are measured as a function of lag-time $\tau = t - t_{w}$ for each $t_w$. 
While the MSD after the second quench differs by a magnitude compared to Fig.~\ref{fig:MSD-T1}, 
the $t_{w}$ dependence shows that rather than continuing the dynamics 
from the prior quench, the process re-initializes and dynamics are refreshed based on 
$t_w$, the age of the system following the second quench at $t_1$. It is apparent from 
this analysis that in the cluster model the intervening quench to $\beta_2$ (if it is not 
excessively long, see below) leaves little mark on the large fraction of frozen-in 
clusters, which on re-heating at $t_1+t_2$ continue their mobility where it froze in at 
$t_1$.

We note that the coarse-grained 
motion in our model \emph{by design} eliminates both, the (trivial) initial 
ballistic motion and the subsequent rattle particles experience at the shortest times 
while confined within their cages. Such in-cage rattle contributes to a visible plateau in the MSD of experiments or of continuum MD simulations, as seen in Figs.~3 and~4(a) of Ref.~[\citenum{Scalliet19}]. Accordingly, such a plateau is absent in our study, in which 
particles are bound to discrete lattice sites until an actual event occurs.

One aspect of rejuvenation in spin glasses~\citep{Bouchaud01} or the MD simulations~\citep{Scalliet19} that the cluster model can not reproduce concerns the infinite time limit $t_1\to\infty$. Even in that case, rejuvenation -- albeit in a very weak form -- is observed. This may not be too surprising, since this situation parallels the original quench that puts a glassy system out of equilibrium into an aging state, only that this quench to $\beta_2>\beta_1$ commences from a temperature \emph{below} the glass transition, $\beta_1>\beta_g$, instead of from $T=\infty$ ($\beta=0$). Nonetheless, the glassy system is dislodged from equilibrium into an non-equilibrium state, however minutely, and aging ensues. In the cluster model, such an equilibrium state does not exist: at infinite time for $\beta_1>\beta_g$, there would be just one large cluster spanning the system, whose eventual break-up would erase all memory of $\beta_1$ and any distinction with $\beta=0$.

\begin{figure}
\includegraphics[width=0.99\columnwidth]{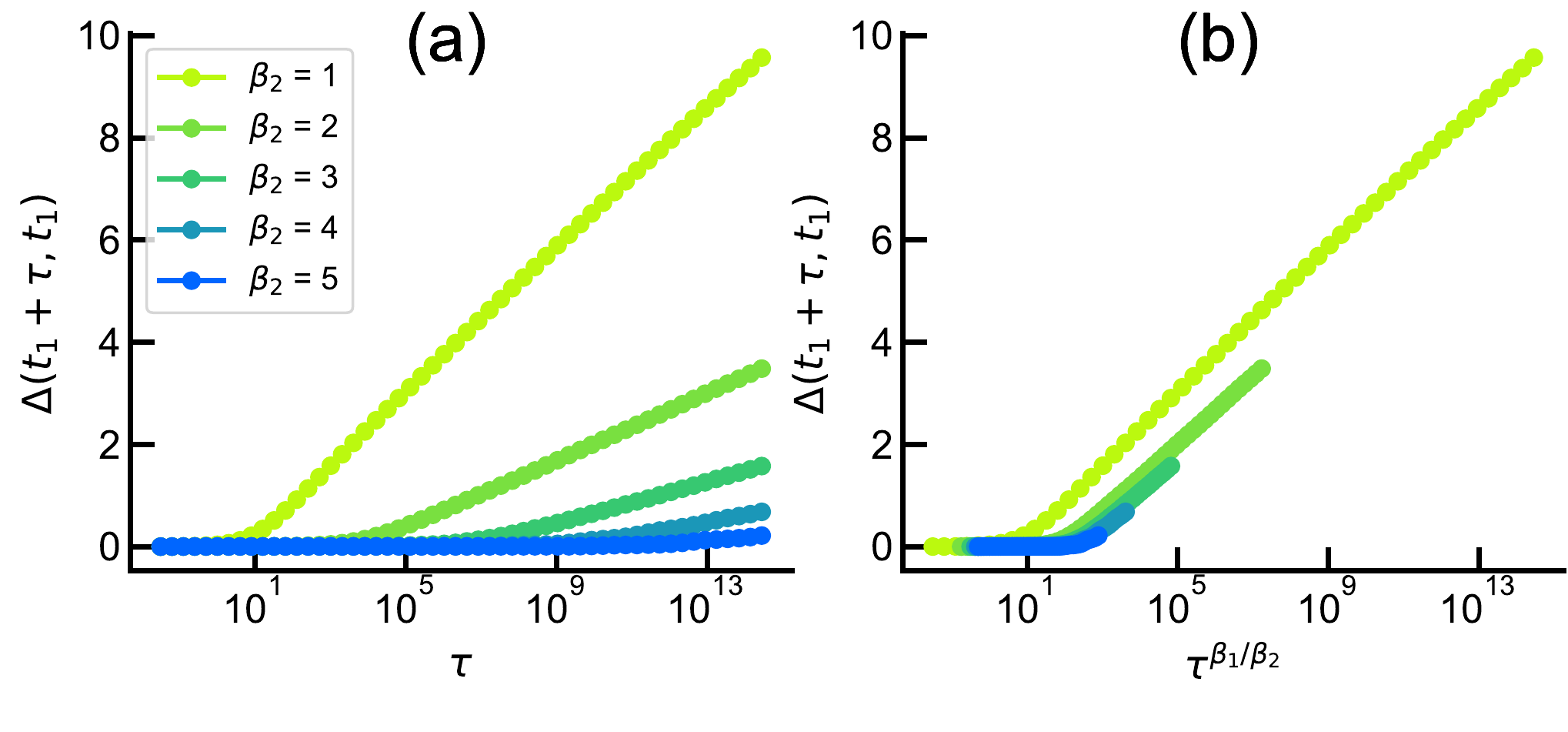}
\caption{\label{fig:prediction}Demonstration for the end of memory. In (a), we measure the MSD $\Delta(t_1+\tau,t_1)$ for particles in the cluster model initially
quenched to $\beta_{1} = 1.0$, then aged for $t_1=25$ sweeps, when it undergoes
the second quench to $\beta_{2}$. The system remains entrenched in its meta-stable state attained at $t_1$ for a time $\tau=\tau_2$ that depends on $\beta_2$, before significant displacement occurs that erases the memory of that state. In (b), this data collapses when $\tau$ is rescaled according to Eq.~(\ref{eq:memory}).
 }
\end{figure}

Finally, we point out that the cluster model reproduces other properties predicted for systems exhibiting rejuvenation and memory effects. For instance, for spin glasses it was shown in Ref.~[\citenum{Bouchaud01}] that the memory effect may diminish for a very long rejuvenation stage. In Fig.~\ref{fig:Rejuvenation-and-memory}, the system ages from the 
initial quench at temperature $\beta_1$ and at $t_1$ has entrenched itself in a 
meta-stable state of some typical free-energy barrier $\Delta F$. To escape the memory of that 
state at $\beta_1$, a record fluctuation is needed, which according to RD typically 
occurs at time  $\tau_1\approx t_0\exp\left\{\beta_1\Delta F\right\}$ with $\tau_1\sim t_1$, where 
$t_0$ is some system-specific microscopic time.  Quenching anew at $t_1$ from $\beta_1$ 
to $\beta_2$ leaves the system even deeper entrenched within \emph{that} state, now 
needing a time $\tau_2\approx t_0\exp\left\{\beta_2\Delta F\right\}$ to escape and lose its 
memory. 
With or without second quench at $t_1$, the clusters formed at $t_1$ with high probability remain stable for $\tau\ll\tau_{1,2}$ but dissolve for  $\tau\gg\tau_{1,2}$, allowing their particles in the process to displace as $\Delta\left(t_1+\tau,t_1\right)\sim A\ln\left(\tau/\tau_{1,2}+1\right)$~\citep{Becker14}.
Thus, if reheating is forestalled until a time $t_2\gg\tau_2$, i.e., a time beyond
 \begin{equation}
     \tau_2\sim B \, t_1^{\beta_2/\beta_1},\label{eq:memory}
 \end{equation}
 with some microscopic constant   $B=t_0^{1-\beta_2/\beta_1}$, memory will have been lost. 
 In Fig.~\ref{fig:prediction}, we demonstrate this effect in the cluster model, evolving the system with the waiting time method over 15 decades.

\section{Conclusions}
\label{conclusions}
In conclusion, we have demonstrated that the cluster model reproduces the 
macroscopic observable rejuvenation behavior of the structural glass studied in 
Ref.~[\citenum{Scalliet19}]. However, we would like to qualify some of their conclusions. 
For one, that their results fit well with predictions of mean-field theory should not 
necessarily be taken as evidence that all aspects of the theory apply to real systems. 
As our dramatically simplified model suggest, an elementary description of the 
requisite hierarchical landscape features~\citep{Robe16,Sibani19} may exist.  That 
mean-field theory also \emph{happens} to provide the same elementary features does not imply that all 
aspects of that theory apply. Rejuvenation and memory by themselves are not even 
sufficient to imply glassy behavior~\citep{Krzakala2006}.

Furthermore, we disagree with the conclusion, based on the pdf of particle 
displacements, ${\cal P}\left(\log\Delta r^2\right)$, shown Fig.~7 in 
Ref.~[\citenum{Scalliet19}], that "all particles are involved in the aging dynamics" 
which occurs "due to very collective particle motion involving the entire system...". 
We observe that these pdf are each distributed around the respective 
plateau values of MSD in Figs.~3 and~4(a) of Ref.~[\citenum{Scalliet19}], thus merely
representing ordinary in-cage rattle. While this may appear as "featureless" and 
therefore homogeneous, it has been shown that the actual \emph{irreversible} events that drive 
relaxation during aging are highly intermittent and localized~\citep{Sibani05}, and are 
likely hidden deep within the large-$\Delta r^2$ tail of those pdf. (Note, e.g., the 
minute bump near $\Delta r^2\approx10^0$ in Fig.~7 of Ref.~[\citenum{Scalliet19}].) Ultimately, this heterogeneity is exactly what 
is captured by the break-up of clusters in our model, after coarse-graining out the in-cage rattle, as that rattle only rarely amounts to meaningful (record-sized, irreversible)  displacements ~\citep{BGS21}.

\section*{Conflicts of interest}
There are no conflicts to declare.

\section*{Acknowledgements}
We thank Paolo Sibani and Eric Weeks for many enlightening discussions.

\section*{References}
\bibliographystyle{vancouver}
\bibliography{/Users/sboettc/Boettcher.bib}

\end{document}